\documentclass[prl,reprint,twocolumn,preprintnumbers,amsmath,amssymb]{revtex4}  

\usepackage{graphicx}
\usepackage{dcolumn}
\usepackage{bm}
\usepackage{multirow}
\usepackage{color}
\usepackage{mathtools}
\usepackage{amsmath}

\DeclarePairedDelimiter\bra{\langle}{\rvert}
\DeclarePairedDelimiter\ket{\lvert}{\rangle}
\DeclarePairedDelimiterX\braket[2]{\langle}{\rangle}{#1 \delimsize\vert #2}
\DeclarePairedDelimiterX\inner[2]{\langle}{\rangle}{#1,#2}

\begin{document}
\title{Dependence of the hBN Layer Thickness on the Band Structure and Exciton Properties of Encapsulated WSe$_2$ Monolayers}

\author{Iann C. Gerber}
\email{igerber@insa-toulouse.fr}
\author{Xavier Marie}
\affiliation{LPCNO, Universit\'e F\'ed\'erale de Toulouse Midi-Pyr\'en\'ees, INSA-CNRS-UPS,135 Av. de Rangueil, 31077 Toulouse, France}


\begin{abstract}
The optical properties of two-dimensional transition metal dichalcogenide monolayers such as MoS$_2$ or WSe$_2$ are dominated by excitons, Coulomb bound electron-hole pairs. Screening effects due the presence of  hexagonal-BN surrounding layers have been investigated by solving the Bethe Salpeter Equation on top of \textit{GW} wave functions in density functional theory calculations. We have calculated the dependence of both the quasi-particle gap and the binding energy  of the neutral exciton ground state E$_b$ as a function of the hBN layer thickness. This study demonstrates that the effects of screening at this level of theory are more short-ranged that it is widely believed. The encapsulation of a WSe$_2$ monolayer by three sheets of hBN ($\sim$ 1 nm) already yields a  20 \% decrease of E$_b$ whereas the maximal reduction is $ 27 \%$ for thick hBN.  We have performed similar calculations in the case of a WSe$_2$ monolayer deposited on stacked hBN layers. These results are compared to the recently proposed Quantum Electrostatic Heterostructure approach.
\end{abstract}

\maketitle

\textit{Introduction.---}Monolayers of Transition Metal Dichalcogenides (TMDCs), based on Mo or W metals have been the subject of intense research thanks to a fascinating combination of properties \cite{Wang:2018cl}. Indeed, TMDC monolayers exhibit a direct fundamental band-gap in the optical range \cite{Mak:2010do,Splendiani:2010hw} contrary to graphene, which makes them good materials for a new generation of optoelectronic devices~\cite{Wang:2012fa}. 
Interestingly, the electronic and optical properties of these systems are dominated by robust excitons, i.e Coulomb bound electron-hole pairs~\cite{Chernikov:2014ic,Ye:2014hb,He:2014bo,Wang:2015kb}. Their very large binding energies of few hundred meV are due to a reduced Coulomb screening in combination to a strong 2D quantum confinement and a large effective mass. Owing to the extreme confinement  in the perpendicular direction to the TMDC plane, excitons are particularly sensitive to the local environment surrounding the monolayer either because of the presence of a substrate  or even more complex stacking configurations~\cite{Lin:2014jw,Borghardt:2017gy,Gupta:2017cl,Florian:2018kk,Park:2018hx,Vaclavkova:2018gs,Hu:2018hw}. In general, the band structures of the constituent 2D crystals and band alignment across the interfaces of a van der Waals Heterostructure (vdWH) depend on many factors: interlayer hybridization, charge transfer, dielectric screening, proximity induced spin- orbit interactions, etc. However in many vdWHs  based on TMDCs, the weak interlayer binding suggests that each vdWH individual layers mainly keep their original 2D properties modified only by the long-range Coulomb interaction with their immediate neighboring layers. 
From an experimental point of view, one usually observes a rather weak dependence of the exciton ground state absolute energy by varying the dielectric environment since both the quasi-particle gap (also called free carriers gap, noted E$_g$ below)  and the exciton binding energy vary and the corresponding changes almost compensate each other. However, recent measurements of the diamagnetic shift of the exciton transition in high magnetic fields evidenced clearly the change of exciton size and binding energy by  encapsulating WSe$_2$ monolayers with hBN \cite{Stier:2016cv,Zipfel:2018jk}. A significant reduction of the exciton binding energy has also been observed by encapsulation of TMDC monolayers with graphene layers~\cite{Raja:2017cxa}. Using metal such as Au as substrate has been also reported, showing a large reduction of the neutral exciton binding energy~\cite{Park:2018hx}.

 The  dependence of the TMDC monolayer band structure and exciton binding energy as a function of the substrate thickness has been less studied~\cite{Ugeda:2014cy,Wang:2017fm, Raja:2017cxa}. The key fundamental question is : what is the typical range for the dielectric environment influence ?
The variation of the exciton energy of WS$_2$ monolayer as a function of the number of layers of capping graphene was measured \cite{Raja:2017cxa}; the impact of a single layer of graphene was clearly evidenced.
The encapsulation with hBN, a material with weaker dielectric constant compared to graphene, presents major advantages as it yields very narrow, homogeneously broadened exciton lines which allow for instance the realization of atomically thin mirrors~\cite{Cadiz:2017ik, Manca:2017jt, Ajayi:2017bv, Robert:2018kb, Scuri:2018ft, Back:2018jj}. In comparison with WSe$_2$ ML deposited on SiO$_2$ substrate, redshifts of the neutral exciton energy in the range 15-30 meV were observed when the ML is transferred on hBN thick layer substrate or encapsulated by hBN~\cite{Raja:2017cxa,Cadiz:2017ik,Borghardt:2017gy}.
 
On a theoretical level, the effect of an  anisotropic dielectric environment on the excitonic properties of an atomically thin layer is a complex issue with non-analytic solutions. A first approach is to use  an approximate form for the radial dependence of  Coulomb potential in the framework of the Wannier-Mott Hamiltonian for thin films~\cite{Rytova:ta, Keldysh:ta, Cudazzo:2011iz, Zipfel:2018jk, 2018arXiv180100477V}. More recently several models have been proposed to study the evolution of the Quasi-Particle band-gaps and the excitonic properties from the ML to the bulk limit \cite{Meckbach:2018hw,Cho:2018fya}; much fewer studies investigate the effects of the substrate dielectric environment  \cite{Latini:2015fl,Kylanpaa:2015cn,Trolle:2017bv,Steinhoff:2018ev} or more complex heterostructures \cite{Rosner:2016jo,Klots:2018jd,Cavalcante:2018ft,Florian:2018kk}. Among these approaches, the Quantum Electrostatic Heterostructure (QEH) model is probably the most popular to study complex vdWHs \cite{Andersen:2015he}, due to its versatility and simplicity \cite{Latini:2017gk}. 

Very few {\textit{ab-initio}} studies at the Density Functional Theory (DFT) level and beyond can be found in the literature. Indeed, the incommensurate nature of vdWHs presents a great challenge for first-principles calculations because it is generally not possible to represent the heterostructure in a computational cell that is small enough to allow the calculation to be performed without straining one or more of the layers and thereby alter its electronic properties. The problem is particularly severe for many-body calculations for which the computational cost grows rapidly with system size. However for some specific and well defined configurations~\cite{Koda:2016jb}, $GW$ coupled to Bethe Salpeter Equation (BSE) calculations ~\cite{Hedin:1965tu,Hanke:1979to,Rohlfing:1998vb} become computationally tractable, allowing the determination of both the QP gap and exciton ground state energy. The most significant works are mainly devoted to investigate the dielectric screening that affects excitons in TMDCs in the vicinity of graphene or graphite thick substrate. In these cases, large band-gap renormalization, of the order of 100 meV, together with a reduction of the exciton binding energy of the same order of magnitude were calculated ~\cite{Ugeda:2014cy,Winther:2017kv}. By using high-k dielectrics environment such as Au metal substrate, even larger band gap reduction can be achieved for MoS$_2$~\cite{Ryou:2016km} or MoSe$_2$ ML\cite{Gupta:2017cl}.  For MoS$_2$ ML on a hBN monolayer  a band-gap reduction of 40 meV (compared to free-standing ML)  has been reported when a simplified model of bulk hBN is used~\cite{Naik:2018bo}, whereas a previous calculations with bulk hBN substrate predicted a  160 meV reduction~\cite{Druppel:2017hs}.  A small redshift of the neutral exciton peak (usually denoted A:1s), around 20 meV,  has also been reported \cite{Druppel:2017hs}, without studying hBN thickness effects.  By increasing the capping layer thickness, the change in band-gap is expected to occur on ultra-short length scale since the main contribution to the self energy are the non-local inter-orbital exchange terms, which directly control the hybridization and the resulting electronic band gap. Since those non-local contributions are mainly localized within a radius of about three units cells \cite{Rosner:2016jo}, we can expect to mimic thick substrate and encapsulation effects by using just a few hBN layers. However, a systematic calculation of the band structure and exciton properties of a TMDC monolayer as a function of the hBN surrounding layer thickness is still lacking. The knowledge of this dependence is crucial for the engineering of vdWHs.

\begin{figure}[htp]
\includegraphics[width=0.45\textwidth]{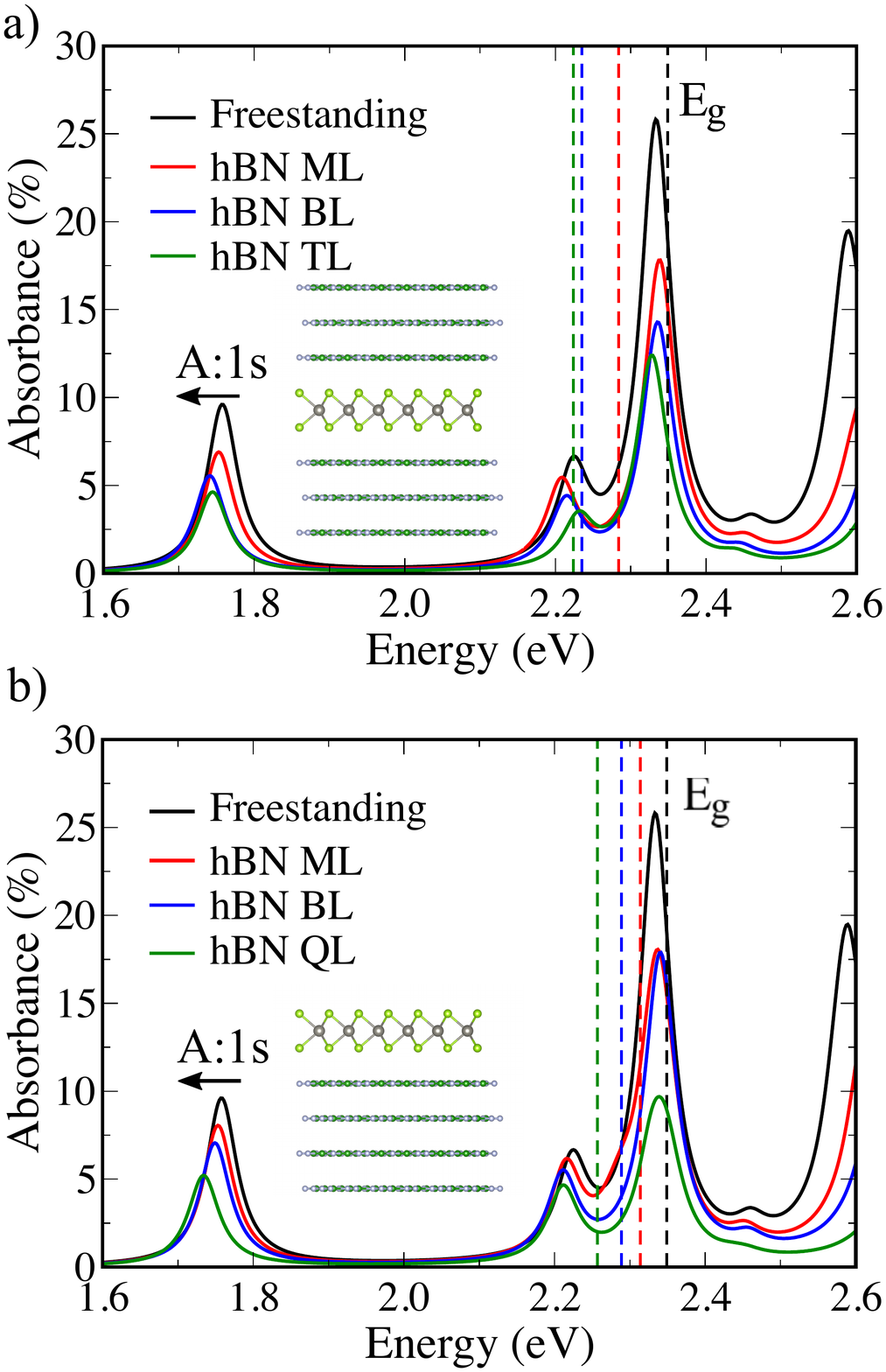}
\caption{\label{fig:absorb} a) Absorbance spectra of a WSe$_2$ monolayer encapsulated by h-BN mono-, bi- or tri-layers b) Absorbance spectra of a WSe$_2$ monolayer stacked on a h-BN mono-, bi- or quadri-layers. The corresponding fundamental band-gaps values are given by dashed lines, when the insets recall the stacking geometries: W and Se are in grey and light green respectively, when B and N are in light blue and green.}
\end{figure}

\begin{figure}[htp]
\includegraphics[width=0.45\textwidth]{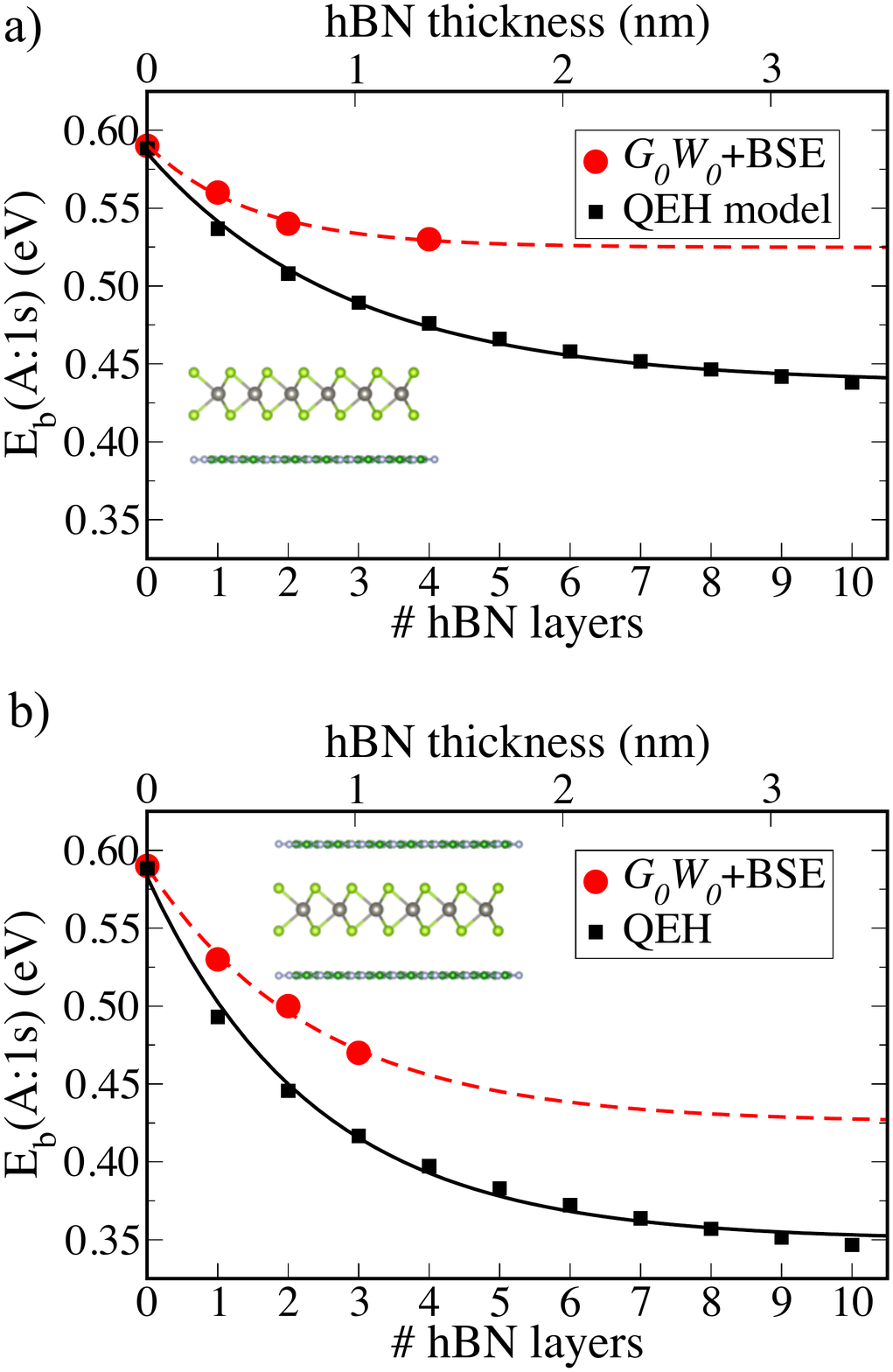}
\caption{\label{fig:Ebind} Evolution of the A:1s exciton binding energy with respect to hBN thickness, using $G_0W_0$+BSE and QEH model of Ref.\cite{Latini:2015fl} a) when hBN is used as substrate and b) when  hBN is used for encapsulation. Bold and dashed lines correspond to interpolated curves using an exponential decay curves. The insets recall the stacking geometries: W and Se are in grey and light green respectively, when B and N are in light blue and green.}
\end{figure}

In this paper we have calculated the dependence of both the free carrier gap E$_g$ and the binding energy E$_b$ of the neutral exciton ground state  as a function of the hBN layer thickness. As shown in Figure~\ref{fig:absorb}, we demonstrate, thanks to $GW$+BSE approach that both E$_g$ and E$_b$  are efficiently tuned by using insulating hBN encapsulation layer due to environment dielectric screening. For thick hBN  ( $>$ 10 MLs), we find a decrease of the exciton binding energy by about 27\% , i.e. 160 meV (compared to free-standing ML), as extrapolated in Figure \ref{fig:Ebind}. These results are in rather good agreement with recent measurements \cite{Raja:2017cxa, Cadiz:2017ik, Borghardt:2017gy, Stier:2018kg}.
The striking feature is that the encapsulation of the WSe$_2$ monolayer by only three sheets of hBN  ($\sim$ 1 nm) already yields a  20 \% reduction of the exciton binding energy. As expected smaller binding energy reduction occurs when hBN is only used as a substrate. We also show that the QEH model tends to overestimate the reduction of the binding energies when compared to our first-principles calculations in both stacking configurations.     

\begin{figure}[htp]
\includegraphics[width=0.45\textwidth]{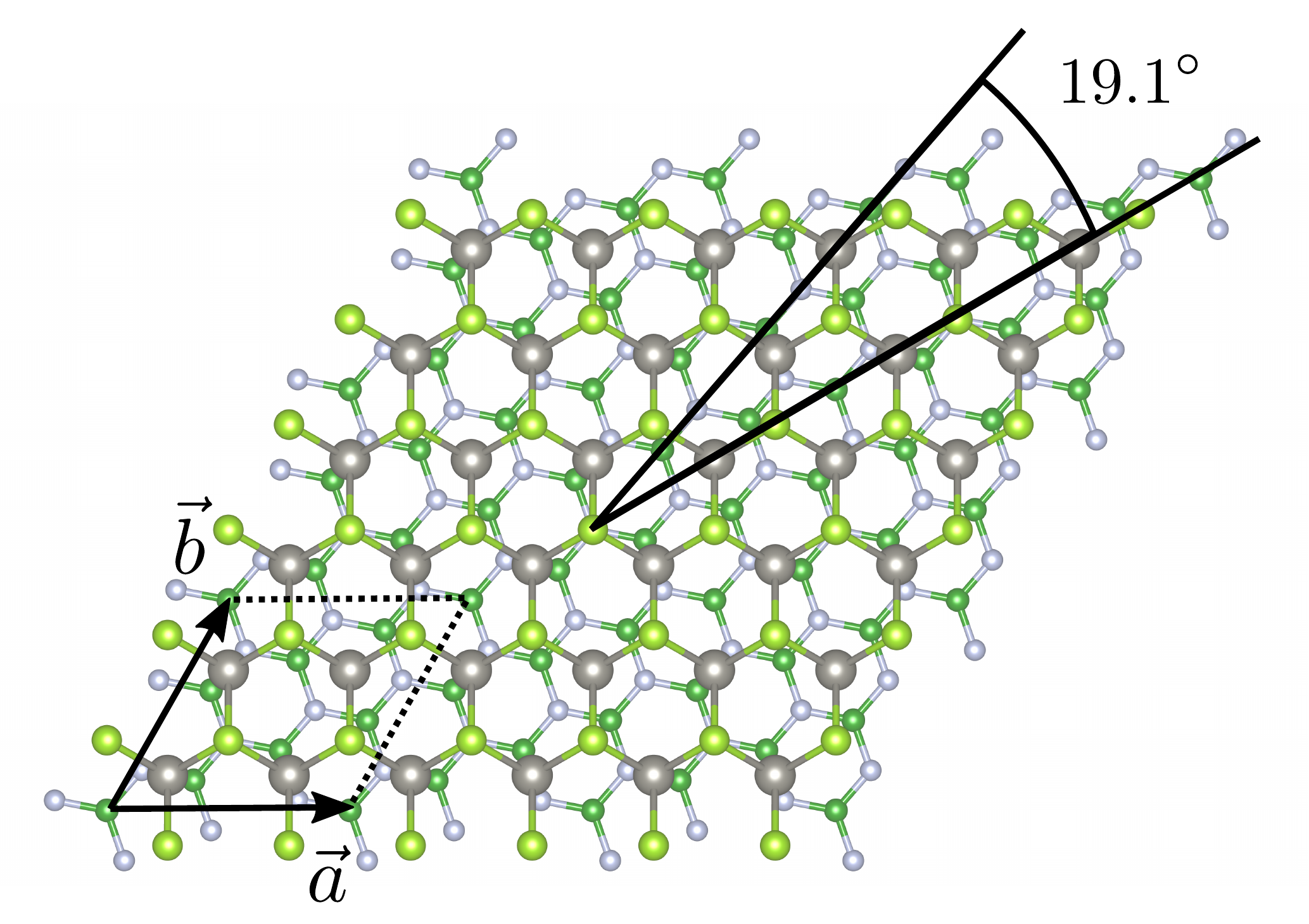}
\caption{\label{fig:stacking} Schematic of the hBN/WSe$_2$ lattice structure presenting the calculation cell ($\vec{a}$, $\vec{b}$), with a 19.1$^ \circ$ rotation between both lattices, W and Se are in grey and light green respectively, when B and N are in light blue and green.}
\end{figure}

\indent \textit{Computational Details.---} 
The atomic structures, the quasi-particle band structures and optical spectra are obtained from DFT calculations using the VASP 
package \cite{Kresse:1993a,Kresse:1996a}. It uses the plane-augmented wave scheme \cite{blochl:prb:94,kresse:prb:99} to treat core electrons. 
Three, five, six and fourteen electrons have been explicitly included in the valence states for B, N, Se and W respectively.
Perdew-Burke-Ernzerhof (PBE) functional \cite{Perdew:1996a} is used as an approximation of the exchange-correlation electronic term, to build the wavefunction, which serves as starting point for further $G_0W_0$ calculations.  
During geometry's optimization step of all the hetero-structures, performed at the PBE-D3 level~\cite{Grimme:2010ij}, all the atoms were allowed to relax with a force convergence criterion below $0.005$ eV/\AA,  in order to include van der Waals interaction between layers. The optimized lattice parameter of WSe$_2$ used for all the calculations is 3.32 \AA. The coincidence lattice method for 2D crystals, as proposed in Ref.~\cite{Koda:2016jb}, has been used to generate computationally tractable supercells,    
with the aim of also minimizing the strain on h-BN layers. Thanks to the \textit{CellMatch} software \cite{Lazic:2015ke} we have generated the h-BN/WSe$_2$ supercell with the following parameters:  a ($\sqrt{7} \times \sqrt{7})$ R 19.1$^{\circ}$/  (2 $\times$ 2), as shown in Fig.~\ref{fig:stacking}, which corresponds to biaxial strain of 0.5\% on the h-BN layers. Within this geometry two stacking ordering are available: either the Se atom in (1/3, 1/3)  position is lying above a B or N atom. In the case of a single hBN ML, the energy difference between the two stackings is less than 1 meV, when the interlayer distance (d) are 3.42 and  3.48 \AA~for B-aligned and N-aligned respectively. When including more h-BN layers  we have used AA'-stacked geometry yielding a hBN-hBN interlayer distance of 3.39 \AA since it appears that the eclipsed configuration is the most stable one for hBN bulk and bilayers,\cite{Constantinescu:2013bh}. Effective Band Structures (EBS) on top of PBE calculations have been calculated using unfolding technics proposed in ref.~\cite{Medeiros:2014ka,Medeiros:2015ks}.    

 A grid of 6$\times$6$\times$1 k-points has been used, in conjunction with a vacuum height of 18.4 \AA, for all the calculation cells, to take benefit of error's cancellation in the band gap estimates~\cite{Huser:2013fs}, 
and to provide absorption spectra in good agreement with experiments~\cite{Klots:2014a, MolinaSanchez:2013a}. An energy cutoff of 400 eV and a gaussian smearing of 0.05 eV width have been chosen for partial occupancies, when a tight electronic minimization tolerance of $10^{-8}$ eV is set to determine with a good precision the corresponding derivative of the orbitals with respect to $k$ needed in quasi-particle band structure calculations. 
Spin-Orbit Coupling (SOC) was also included non-self-consistently to determine eigenvalues and wave functions as input for the full-frequency-dependent $GW$ 
calculations~\cite{Shishkin:2006a} performed at the $G_0W_0$ level ~\cite{Echeverry:2016ev}. The total number of states included in the $GW$ procedure is set to 960, in conjunction with an energy cutoff of 100 eV for the response function, after a careful check of the direct band gap convergence (smaller than 0.1 eV as a function of k-points sampling).
 All optical excitonic transitions have been calculated by solving the Bethe-Salpeter Equation as follows~\cite{Hanke:1979to,Rohlfing:1998vb}:
\begin{equation}
\left( \varepsilon_c^{\textrm{QP}}-\varepsilon_v^{\textrm{QP}}\right) A_{vc} + \sum_{v'c'} \bra{vc}K^{eh}\ket{v'c'}A_{v'c'}=\Omega A_{vc},
\end{equation}
where $\Omega$ are the resulting $e$-$h$ excitation energies. $A_{vc}$ are the corresponding eigenvectors, when $\varepsilon^{QP}$ are the single-quasiparticle energies obtained at the $G_0W_0$ level, and $K^{eh}$ being the CB electron-VB hole interaction kernel.  
This term consists in a first attractive screened direct term and a repulsive exchange part. Practically we have included the eight highest valence bands and the twelve lowest conduction bands to obtain eigenvalues and oscillator strengths on all systems. From these calculations, we report the absorbance values by using the imaginary part of the complex dielectric function $\epsilon_2(\omega)$, with the following formula  ~\cite{Yang:2009dp}:
\begin{equation}\label{eq:abs}
A(\omega)=\frac{\omega}{c}\epsilon_2(\omega)\Delta z,
\end{equation}
where $\Delta z$ is the vacuum distance between periodic images, thus this quantity  should not depend on the size of calculation cells in the perpendicular direction. As pointed out by Bernardi \textit{et al}~\cite{Bernardi:2013ie}, Eq.(\ref{eq:abs}) is a Taylor expansion for $\Delta z \to 0$ of the absorbance defined as $\displaystyle A= 1-\textrm{e}^{-\alpha_2 \Delta z}$ for a single or bi-layer of a bulk material with a thickness $\Delta z$, presenting an absorption coefficient $\displaystyle \alpha_2(\omega)=\frac{\omega~\epsilon_2(\omega)}{c~n(\omega)}$ ; here the refractive index is $n=1$, since the considered hetero-structure is surrounded by vacuum only. The computational setup used to extract the binding energies out of the QEH model in Figure~\ref{fig:Ebind} is the following: the hBN-hBN and WSe$_2$-hBN distances are 3.4 \AA~and 4.7 \AA~respectively as in DFT calculations. An effective mass of 0.29 m$_\textrm{0}$ is used to yield the same binding energy of the A:1s exciton as in our $G_0W_0$+BSE calculations.

\begin{table}[htp]
\caption{\label{tab:gap} Calculated WSe$_2$ monolayer quasi-particle band-gaps from DFT and $G_0W_0$ and A and B exciton peak position in eV, upon encapsulation with hBN layers (upper part of the table) or substrate modeling .}
\begin{center}
\begin{tabular}{ccccccccc }
\hline
\hline
& & DFT & & $G_0W_0$ & & A & & B\\
\hline
Freestanding & & 1.28 & & 2.35 & & 1.76  & & 2.22 \\
hBN ML-encapsulation & & 1.25 & & 2.28 & & 1.75 & & 2.21 \\
hBN BL-encapsulation  & & 1.25 & & 2.24 & & 1.74 & & 2.21 \\
hBN TL-encapsulation & & 1.25  & & 2.22 & & 1.75 & & 2.23 \\
\hline
hBN ML-substrate & & 1.25 & & 2.30 & & 1.74 & & 2.20 \\
hBN BL-substrate & &1.25  & & 2.29 & & 1.75  & & 2.21 \\
hBN QL-substrate & & 1.25  & & 2.26 & & 1.73 & & 2.21 \\
\hline
\hline
\end{tabular}
\end{center}
\end{table}

\indent \textit{Bang gap variations with the dielectric environment.---} Absorption spectra of an encapsulated WSe$_2$ ML with three different hBN thicknesses,  ML, Bi-Layer (BL) and Tri-Layer (TL)  are presented and compared to the ideal freestanding configuration in Figure~\ref{fig:absorb}(a). Our $GW$ calculations performed on the freestanding ML exhibit a direct QP band gap at the K point, with a value of  2.35 eV and a ground state exciton energy ("optical gap") of 1.76 eV. Consequently, the binding energy of the lowest-energy exciton ((A:1s)  is 0.59 eV. These two values are in line with previous theoretical and recent experimental studies~\cite{Ramasubramaniam:2012eb, Echeverry:2016ev, Borghardt:2017gy,Wang:2015kb}. Table \ref{tab:gap} summarizes QP (calculated at the $G_0W_0$ level)  and DFT band gaps, as well as A and B ground state exciton peak positions, for different dielectric environments. If the surrounding dielectric environment screens efficiently the electron-hole interaction by decreasing the A peak positions and at the same time reduces the fundamental band gap, it also leaves the A-B splitting unchanged. This confirms that this splitting is solely due to SOC and it remains largely unaffected by the presence of any dielectric environment. 

Upon encapsulation we observe a decrease of the QP band gap by already 110 meV when hBN BLs are used, when it becomes 130 meV for TLs (Figure~\ref{fig:absorb}(a)). This means that very thin hBN environment has a significant effect on the screening of the interactions within the TMDC sheet. We have also investigated the energy changes if hBN lies only on one side of the TMDC ML ((Figure~\ref{fig:absorb}(b))). When the WSe$_2$ is stacked on a hBN Quadri-Layer (QL) the corresponding QP band gap is also significantly reduced by 90 meV. This value is in good agreement with the recent measurement of the QP band gap change ($\sim$ 100 meV) between WSe$_2$ ML deposited on a 8nm hBN layer and WSe$_2$ ML deposited on a thick SiO$_2$~\cite{Raja:2017cxa}. Moreover in Ref.~\cite{Naik:2018bo},  a value of 40 meV QP energy change was reported in the case of MoS$_2$ ML on hBN whereas it is extrapolated to be 160 meV in the work of Dr\"uppel {\textit{et al.}}~\cite{Druppel:2017hs} Those results stress the importance of using post-DFT approaches that account for many-body term corrections to the self-energy of electrons
in such vdWH structures. Indeed Table \ref{tab:gap} shows that the band gap reduction is much smaller (30 meV) for the same systems at the PBE level of theory, and there is no distinction between the stacking or encapsulation situations. More importantly the independent particle band gap value already saturates even in the case of the TMDC ML in interaction with a single hBN layer, see Table \ref{tab:gap}. 

\begin{figure}[htp]
\includegraphics[width=0.45\textwidth]{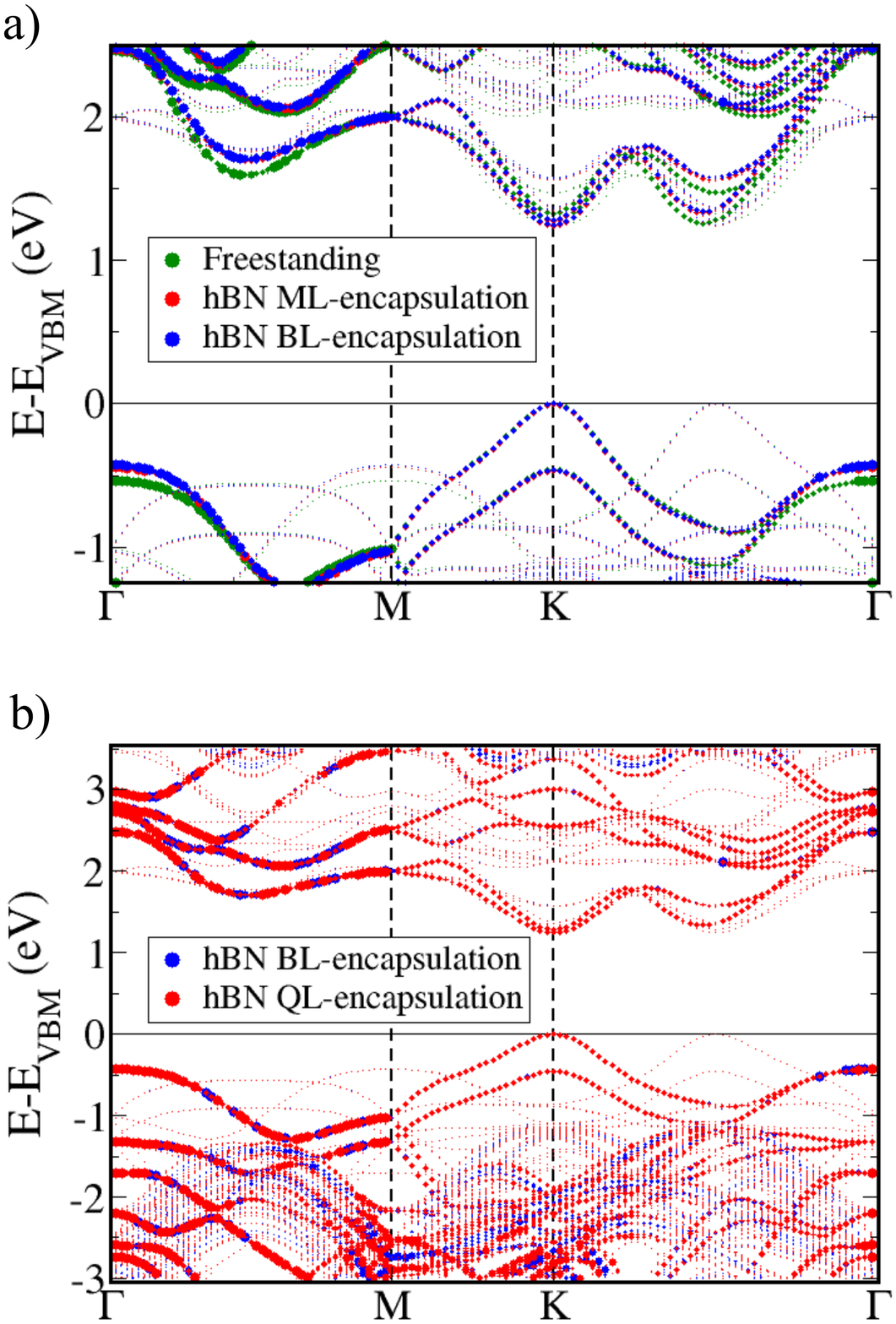}
\caption{\label{fig:unfold} a) Evolution of the Effective Band Structure at the PBE level, upon encapsulation with hBN layers of a  WSe$_2$ ML, with 1ML of hBN (red) on each side or 1BL of hBN on each side (blue). Mind that the BL and ML encapsulation situation are almost super-imposed over the entire first-Brillouin zone. 
b) Comparison of the Effective Band Structure of a WSe$_2$ ML encapsulated by hBN BLs and hBN QLs, showing that the PBE level of calculations misses long-range electronic screening, since the two band structures are super-imposed over the entire first-Brillouin zone.}
\end{figure}

The limitation of standard DFT to investigate electronic properties of vdWHs is also clearly shown in Figure~\ref{fig:unfold}, where no differences in the Effective Band Structure of  hBN BL/WSe$_2$/hBN BL and hBN QL/WSe$_2$/hBN QL stacking are visible. Interestingly when comparing with the effective band structure of a freestanding ML, the presence of the surrounding hBN layers environment always pushes upward the valence bond maxima in $\Gamma$, as well as in the conduction band mimima in the valley (Q) located between K and $\Gamma$ points. The origins of the too small band gaps values and the limited effects due to the presence of hBN layers for this level of calculation are (i) the lack of self-energy correction in a standard PBE type of calculations and (ii) the exponential decay of the exchange term, which is essentially based on a Slater-Dirac expression which roughly behaves as $\rho^{4/3}$ where $\rho$ stands for the electronic density, despite gradient corrections~\cite{Becke:1988tx}.  As a consequence, at the standard DFT level, the presence of hBN layers does not significantly change the calculated band gap value, since the dielectric environment is enable to affect the TMDC intra-layer hybridization between the $d$ orbitals of the TM and the $p$ orbitals of the chalcogens, which controls the gap opening in the K valley between the $d_{z^2}$ and $d_{xy}, d_{x^2-y^2}$ states~\cite{Mattheiss:1973tz}. This limitation of standard DFT calculations has been recently reported in the same context of varying the dielectric environment of TMDC MLs~\cite{Naik:2018bo}.

\indent \textit{Exciton energy variations with the dielectric environment.---}
Additionally to the reduction of the band gap, one can observe in Figure~\ref{fig:absorb}(a)  a slight redshift of the A:1s exciton peak  when the hBN encapsulation layer thickness increases. Upon a hBN TL encapsulation  the calculated shift towards smaller energy is around 20 meV, compared to the freestanding ML. With a hBN QL serving as a substrate model we obtain a redshift of roughly 30 meV, Figure~\ref{fig:absorb}(b). 
As expected these shifts remain much smaller compared to the band gap renormalization due to the simultaneous reduction of the exciton binding energies of the A:1s exciton. Our calculated reduction of exciton energies are in agreement with recent measurements \cite{Raja:2017cxa,Cadiz:2017ik,Borghardt:2017gy}. For instance, Borghardt \textit{et al.}~\cite{Borghardt:2017gy} reported redshifts of the neutral exciton emission energies from $\mu$-photoluminescence experiments when hBN was used as a substrate (17 meV) or for encapsulation (35 meV). This trend was also evidenced in a recent theoretical work on excitons and trions in MoS$_2$ ML  on different substrates~\cite{Druppel:2017hs}.  

All together our calculations displayed in Figure~\ref{fig:Ebind} show that the fundamental excitonic binding energy reduction using three hBN layers (TL) is 120 meV compared to a freestanding ML. We have used a simple exponential decay law based to extrapolate the A:1s exciton binding energies from the $G_0W_0$+BSE and QEH data. After stacking 10 hBN layers ($\sim$ 3 nm) the binding energy becomes clearly insensitive to the use of additional layers. The maximal reduction of the binding energy of the A:1s exciton is then around 27\%  upon hBN encapsulation. This calculated binding energy reduction is in line with experimental determination of the effect of surrounding dielectric environment  Ref.~\cite{Stier:2016cv,Borghardt:2017gy, Stier:2018kg}. Moreover, we note in Figure~\ref{fig:absorb} a clear decrease of the exciton absorbance upon encapsulation. This is perfectly consistent with the decrease of the exciton oscillator strength resulting from the reduction of its binding energy.
We have compared our results to the ones obtained with the Quantum Electrostatic Heterostructure (QEH) model \cite{Andersen:2015he,Latini:2017gk}. It consists on a semi-classical approach which takes as input the dielectric functions of the individual isolated layers computed fully quantum mechanically at the random phase approximation level of theory and couple them classically via their long-range Coulomb interaction. In order to get the exciton properties, this method which yields the global vdWH dielectric function is combined in a second time to a generalized 2D Mott-Wannier exciton model. Figure~\ref{fig:Ebind} compares the variation of the ground state exciton binding energy with our fully {\textit{ab-initio}} approach and the QEH model using hBN encapsulation an hBN as a substrate. When the hBN thickness increases, similar trends are observed with both calculation methods. However for thick hBN thickness ($\sim$ 10 monolayers, i.e 3 nm), the reduction is much larger, around 40\%, for the QEH model, compared to our calculations ($\sim 27\%$).  On can suspect that, as proposed in Ref.~\cite{Florian:2018kk}, the too strong screening in the QEH model originates from the absence of interlayer gaps in the vdWH building.   
To the best of our knowledge no experimental data on the variation of the exciton binding energy of WSe$_2$ ML as a function of the hBN thickness are available. Nevertheless, in quite similar situation, the measured variation of the exciton energy of WS$_2$ ML as a function of the number of graphene capping layer also shows that (i) the QEH model overestimates the redshift \cite{Raja:2017cxa} and (ii) similarly to our calculation, the effects of screening is very short-ranged : even a single hBN layer already has a very significant impact on the exciton binding energy.

\indent \textit{Conclusion.---}
We have calculated the dependence of both the quasi-particle band gap E$_g$ and the binding energy E$_b$ of the neutral exciton ground state in WSe$_2$ monolayer as a function of the hBN encapsulation layer thickness. Our approach consists in solving the Bethe Salpeter Equation on top of \textit{GW} wave functions built from standard DFT calculations. Our calculations show the large variation of both E$_g$ and E$_b$ as a consequence of the tuning of the monolayer surrounding dielectric environment. The key result is that the encapsulation of the WSe$_2$ monolayer by only three sheets of hBN ($\sim$ 1 nm) already yields a  20\% reduction of the exciton binding energy whereas the maximal reduction for a thick hBN layer is $\sim 27 \%$ . As expected smaller binding energy reduction occurs when hBN is only used as a substrate. We also show that the Quantum Electrostatic Heterostructure model tends to overestimate the reduction of the binding energies when compared to our first-principles calculations in both stacking configurations. These results can be very useful to engineer the exciton properties in new van der Waals Heterostructures. Unfortunately our computational setup does not allow the determination of the effect of dielectric screening on the excited states of the excitons, this point certainly deserves further work in relation with recent experimental work~\cite{Borghardt:2017gy,Robert:2018kb,Stier:2018kg,Zipfel:2018jk}. 

\begin{acknowledgments}
I. C. Gerber thanks the CALMIP initiative for the generous allocation of computational times, through the project  p0812, as well as the GENCI-CINES and GENCI-IDRIS for the grant A004096649. X. Marie acknowledges the Institut Universitaire de France.
\end{acknowledgments}


\end{document}